\documentclass[aps,prl,twocolumn,superscriptaddress,amsmath,amssymb,showpacs,showkeys,floatfix]{revtex4-2}

\usepackage[english]{babel}
\usepackage[utf8]{inputenc}
\usepackage{natbib}
\usepackage{epsfig}
\usepackage{graphicx}
\usepackage{bm}
\usepackage{amsmath}
\usepackage{physics}
\usepackage{xcolor}
\usepackage{amsmath, nccmath, mathtools}
\usepackage{hyperref}
\usepackage{breakurl}

\DeclarePairedDelimiter\floor{\lfloor}{\rfloor}

\begin{document}

\title{Approach to network failure due to intrinsic fluctuations}
\author{Shaunak Roy}%\textsuperscript{*}}
\email{shaunak.roy@students.iiserpune.ac.in}
\affiliation{Indian Institute of Science Education and Research, Pune 411 008, India.}
\author{Vimal Kishore}%\textsuperscript{\dag}}
\email{vimalk@bhu.ac.in}
\affiliation{Department of Physics, Banaras Hindu University, Varanasi 221 005, India.}
\author{M. S. Santhanam}%\textsuperscript{\ddag}}
\email{santh@iiserpune.ac.in}
\affiliation{Indian Institute of Science Education and Research, Pune 411 008, India.}

\date{\today}

\begin{abstract}
We investigate the asymptotic failure of the network arising solely from intrinsic fluctuations in the flux passing through its nodes. By modeling the flux as random walkers and the node failure induced by large fluctuations as extreme events, it is shown that three distinct nodal failure regimes can be identified before a complete network failure takes place. Further, the approach to network failure through these three regimes is shown to be qualitatively similar for square lattice, all-to-all, scale-free, and Erdos-Renyi networks. We obtain approximate analytical description of the approach to failure for an all-to-all network. This network failure scenario is also demonstrated on a real flight transportation network.
 
\end{abstract}

\maketitle
It is well appreciated from our common experience that when networked systems fail, a large segment of population are affected and it unleashes significant economic losses. For instance, power grid failure or blackout \cite{AlbAlbNak2004} is one such example and, across the world, many large geographical regions have encountered power blackouts ranging from a few hours to even days. Heavy traffic jams on road networks, called grid locks, sometimes linger on several days. In all these examples, a functional network, upon the emergence of extreme events, loses its ability to support and facilitate normal flux and dynamics. A network is said to be resilient if it can resist the failures and resume its normal functions after a strong perturbation \cite{KaiLatWit2021}. A study of such network failures and their resilience continues to attract attention in a general setting of complex networks for more than two decades now \cite{Bianconi_2023,PagAie2013,ValSheLa2020,DeyGelPoo2019,HeiFujAih2014,DiaMonMac2023,TamAmb2023}. 

The central question of interest is about how networks fail in response to perturbation, which can arise either due to external attacks or internal reasons \cite{GroBar2022,LiuLiMa2022,Cohen2010}. Most of the earlier works had focussed on external attacks. It is known that scale-free networks are resilient to random attacks, but not against targeted attacks \cite{RekHawBar2000,LaiMotNis2004}. Modeling studies of internet or in terms of its network analogues \cite{CohEreBen2001} have found a similar response to attacks, and this is type of resilience is dubbed ``robust yet fragile'' \cite{DoyAldLi2005}. In networked systems, cascade of node failures can arise in response to failure of one or many nodes \cite{MotLai2002,KorBarTuc2018,ValSheLa2020}. This typically happens in many critical infrastructure systems such as the power distribution network in which blackouts arise from uncontrolled cascading of nodal failures \cite{MotLai2002,CruLatMar2004,CarLynDob2004,YanNisMot2017,AhmDasJyo2023,WitHelKur2022,NesSloFio2020}, in road networks in the form of spreading of traffic snarls \cite{JunEom2023,CheYanQia2023,VivHan2022}, in connected devices and smart grids in which a cascade of undesired state changes can happen \cite{Xing2021}. Hence, an assessment of resilience of networks is useful factor for understanding of network failures \cite{EngTejMor2022}.

While externally induced attacks have been well studied, it is not yet well appreciated that {\it internal} fluctuations within a network can also be a source of instability that may ultimately lead to failure. In real systems, large fluctuations on a node can arise from the intrinsic stochastic nature of the dynamics taking place on a network and fluctuations exceeding the handling capacity of the node. In this work, we focus on  such a scenario and show how failures propagate from a single node to the whole network. We present two key findings; (i)  internal fluctuations, on its own, can lead to complete network failure, and (ii) broadly, the network failure is approached through three distinct regimes independent of the underlying network architecture..

Let us consider a complex network with $N$ nodes, $E$ edges, and $W$ {\it independent} walkers executing random walk dynamics. Focussing on any one of its nodes (say, $i$-th node), the flux through this node is denoted by $w(t)$, the number of walkers at time $t=1,2,\dots$. Then, the distribution of walkers is known to be a Binomial distribution given by
\begin{equation}
g(w)=\left(\begin{array}{l}
W \\
w
\end{array}\right) p^w ~ (1-p)^{W-w},
\end{equation}
where $p=k/2E$ is the stationary occupation probability associated with the node that has degree $k$. Thus, $0 \le w(t) \le W$ would constitute an event at time $t$ for our purposes. Then, an event at time $t=\tilde{t}$ would be designated as an extreme event if $w > \tau$, where $\tau$ is the threshold for identifying an event as extreme. In the presence case, 
$\tau = \langle w \rangle + q ~ \sigma_w$, where $\langle w \rangle$ and $\sigma_w$ are the mean number of walkers and the associated standard deviation, and both depend on degree $k$ of the node in question. For instance, if $q=4$, then extreme events are those events occurring in the tail of $g(w)$ beyond $4\sigma_w$ from the mean. Then, the probability for the occurrence of extreme events, for fixed $N$ and $W$, is given by
\begin{equation}
P(k) =\sum_{w=\floor*{\tau}+1}^W g(w) = I_p\left( \floor*{\tau}+1, W-\floor*{\tau} \right),
\label{eq:eep1}
\end{equation}
where $\floor*{.}$ is the floor function, and $I_p(.,.)$ is the incomplete Beta function {\cite{abramowitz1964}}. For a scale-free network, it is known that \cite{KisSanAmr2011} occurrence probability of extreme events $P(k)$, in an average sense, is large for small $k$, and is small for hubs ($k \gg 1)$.

Usually, extreme events tend to be catastrophic, and the affected nodes take finite time to recover, {\it i.e.}, relaxation is not instantaneous. For instance, traffic jams take a good amount of time to return to normal conditions, or defective transformers in a power distribution network cannot be replaced instantly (except if redundancy is built in to handle such situations). In the formalism leading to Eq. \ref{eq:eep1}, it was implicitly assumed that the affected node $i_{\rm ee}$ can recover instantaneously after an extreme event \cite{KisSanAmr2011, KisSanAmr2012,KisSonSan2013}. This is not true in practice.
To understand the effect of nodal damage due to extreme events, we assume that the affected node $i_{\rm ee}$ cannot support the flux of walkers any more. The walkers left on $i_{\rm ee}$ at the time when extreme event happens are re-distributed to the neighboring nodes. After {\it every step} of random walk dynamics, additional three-step process is implemented on {\sl every node} : (a) check for the occurrence of extreme event, (b) if extreme event had occurred, the affected node is disabled and walkers are redistributed as per the dynamics on the network, here random walk, to its immediate neighbors. This redistribution is done at the next time step to preserve the probability of the process. The affected node will not take part in the dynamics any more, (c) if extreme event did not occur, random walk dynamics is continued as before.
As $t \to \infty$, more and more nodes are disabled due to extreme event occurrences, and the network ultimately loses the ability to support useful dynamics, at which point the network is said to have reached a ``failed state''. In some networks, a failed state might still be able to support trivial dynamics, {\it e.g.}, walk between just two surviving nodes. In this paper, we consider four different networks and study how this state of network failure is approached both as a function of time and as function of capacity-to-load ratio (defined below).

\begin{figure}[t]
\includegraphics[width=0.55\linewidth]{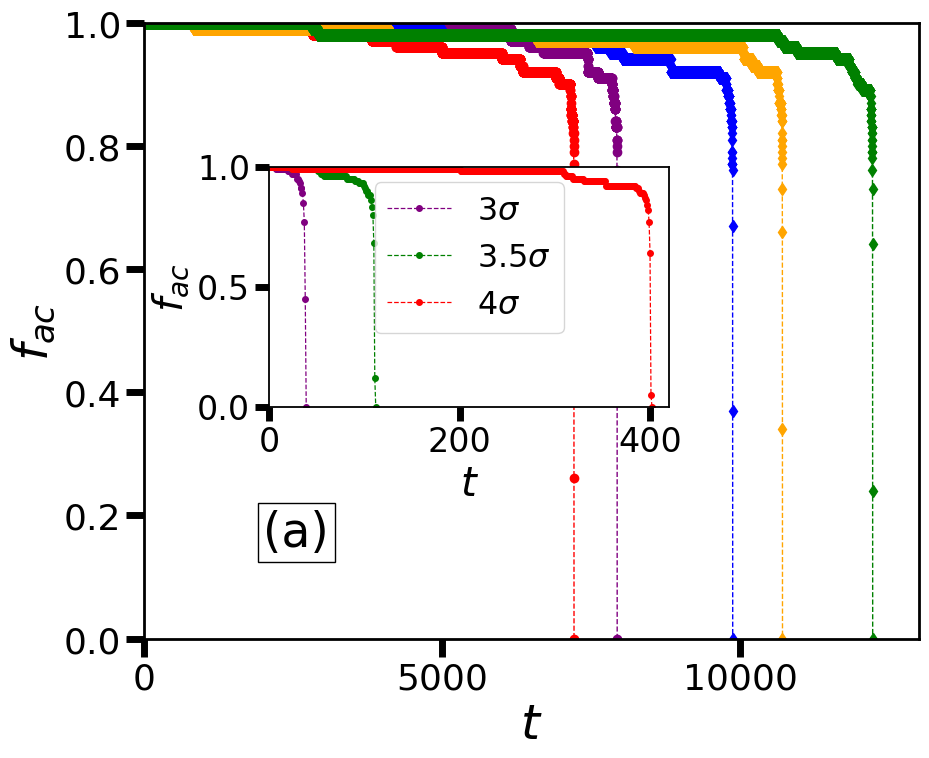}
\includegraphics*[width=0.55\linewidth]{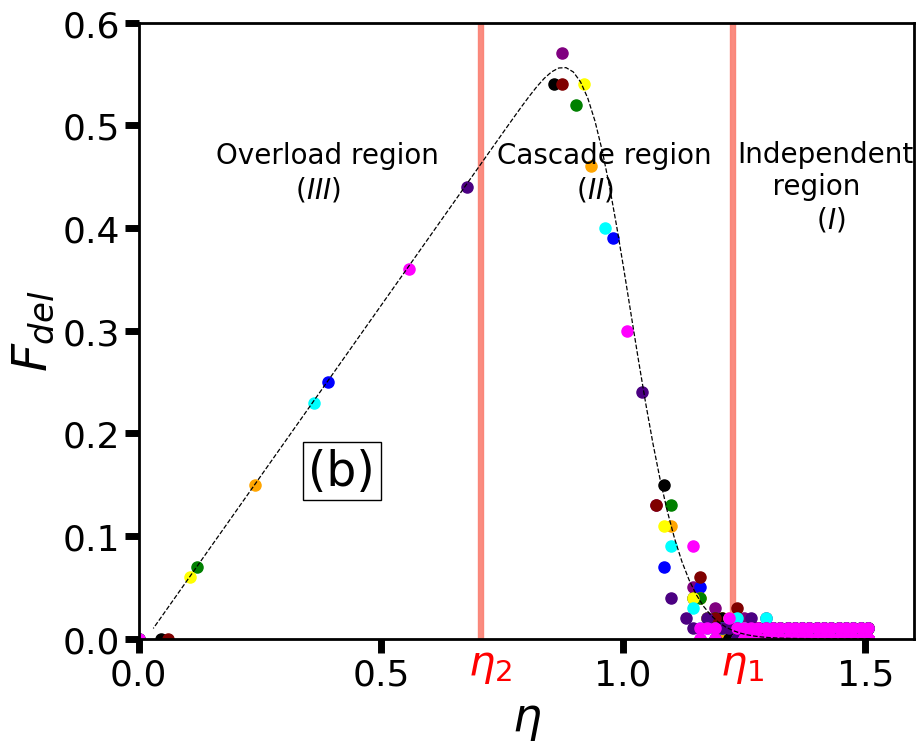}
\caption{Approach to network failure on an all-to-all connected network. (a) fraction of active nodes $f_{\rm ac}(t)$ against time for $5$ different realizations for $q=5$, and (b) instantaneous fraction of deleted nodes $F_{\rm del}(t)$ as a function of $\eta(t)$ (with $ \eta(t=0) = \eta_{0}= 1.5$) for $q=5.0$. The inset in (a) shows $f_{\rm ac}(t)$ for different threshold values $\tau=\langle w \rangle + q \sigma_w$, with $q=3.0, 3.5, 4.0$. The dotted curve in (b) is the analytical result in Eq. \ref{eq:delf1}. Though failure time $T_{\rm fail}$ is different for each realization in (a), all of them collapse on the same (dotted) curve in (b). $\eta_1$ and $\eta_2$ represent boundaries between the regimes (see text for detailed description). }
\label{fig1}
\end{figure}

%\label{sec:a2a}
The central ideas will be illustrated using an all-to-all connected network, and subsequently for other types of networks. Consider an all-to-all network with $N=100$ nodes on which $W=2E=9900$ walkers are executing random walk dynamics. In this work, we have kept the number of walkers $W=2E$ for simplicity. At the initial time, all the walkers are placed randomly on any of the $N$ nodes, and all the nodes begin to send/receive walkers. In our simulations, random initial condition for the walkers is achieved by placing all the walkers initially at an arbitrary node and then performing random walk for sufficiently long time. As random walk dynamics progresses, nodes begin to encounter extreme events. If an extreme event happens on $i$-th node at time $t$, then the node is rendered inactive and stands deleted. Hence, it cannot host random walkers anymore. Let $N_{ac}(t)$ and $N_{del}(t)$ represent, respectively, the cumulative number of active and deleted nodes on the network until time $t$. Then, $f_{ac}(t)=N_{\rm ac}(t)/N$ and $f_{\rm del}(t)=N_{\rm del}(t)/N$ represent the fractions of active and inactive nodes, such that $f_{\rm ac}(t) + f_{\rm del}(t) = 1$. It is useful to define instantaneous number of deleted nodes $n_{\rm del}(t)$ such that $N_{\rm del}(t) = \sum_{t'=1}^t n_{\rm del}(t')$. The corresponding instantaneous fraction is $F_{\rm del}(t) = n_{\rm del}(t)/N$. 

Figure \ref{fig1}(a) displays $f_{\rm ac}(t)$ as a function of time for five  different realization of random walk dynamics. Each realization corresponds to a different initial condition for walkers on the same network. As anticipated, $f_{ac}(t)$ tends monotonically towards zero. Two features can be inferred from this figure; initially, occurrences of extreme events tend to be few and hence $f_a(t)$ is nearly flat. After some time, the number of extreme events increases rapidly and $f_a(t) \to 0$ rapidly. Ultimately, the network fails at $t=T_{\rm fail}$. The failure time $T_{\rm fail}$ is stochastic, and is different for different realizations. Further, as shown in the inset of Fig. \ref{fig1}(a), $T_{\rm fail}$ depends on the value of threshold $\tau$. In general, it is reasonable to anticipate that $T_{\rm fail}$ will be larger for larger $\tau$, though this can be occasionally violated in specific realizations.

A more useful picture for the approach to network failure can be realized by eliminating time as follows. Let $C(t)$ be the capacity of the network defined as the maximum number of random walkers it can carry without breaching the node-dependent threshold $\tau$ on any of its active nodes. The total number of walkers on the network remains constant, but $C(t)$ keeps decreasing with every failure of a node. At the time of network failure, capacity of network to support walker flux becomes significantly small compared to the total number of walkers. This observation motivates us to define a parameter -- capacity to load ratio -- as
\begin{equation}
\eta(t) = \frac{C(t)}{W} = \frac{1}{W} \sum_{i \in \{S_{\rm ac}(t)\} } \tau_i,
\label{eq:eta}
\end{equation}
in which the summation is over the set of active nodes $\{S_{\rm ac}(t)\}$ at time $t$. With temporal evolution, nodes get deleted due to extreme events, and network capacity $C(t)$ decreases, and consequently $\eta(t) \to 0$. In Fig. \ref{fig1}(b), instantaneous fraction of deleted nodes $F_{\rm del}(t)$ is shown as a function of $\eta(t)$ for several realizations. Unlike the temporal perspective in Fig. \ref{fig1}(a), remarkably, all the realizations collapse on the same (dotted) curve. Next, we obtain an approximate analytical description.

\begin{figure*}[t]
\includegraphics[width=0.32\textwidth]{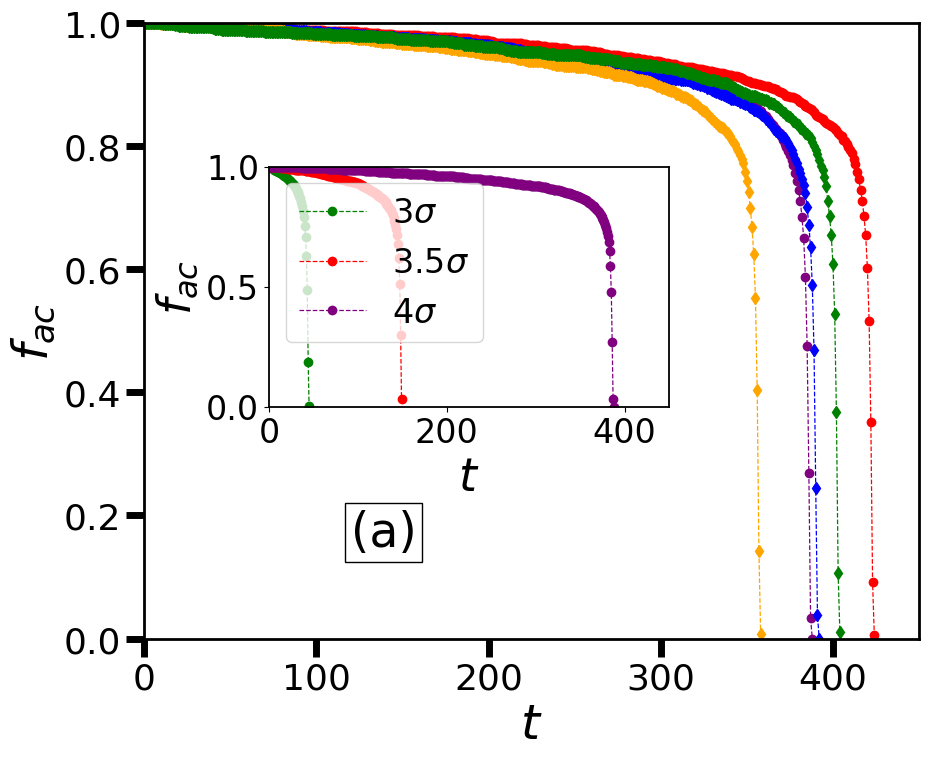}\hspace{0.02\textwidth}
\includegraphics[width=0.32\textwidth]{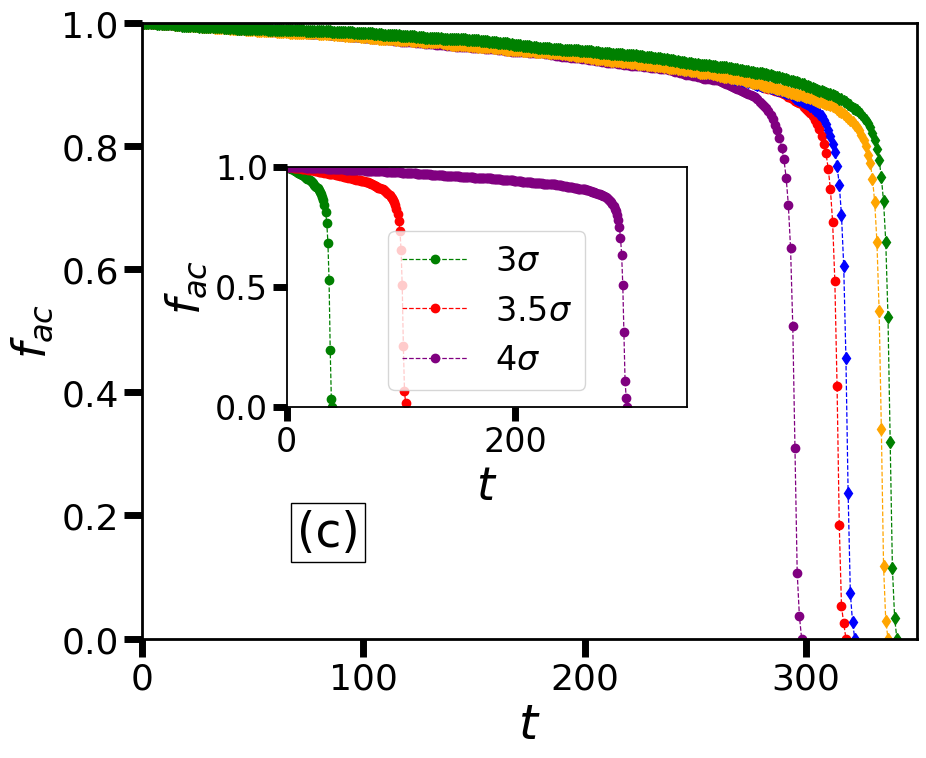}
\includegraphics[width=0.32\textwidth]{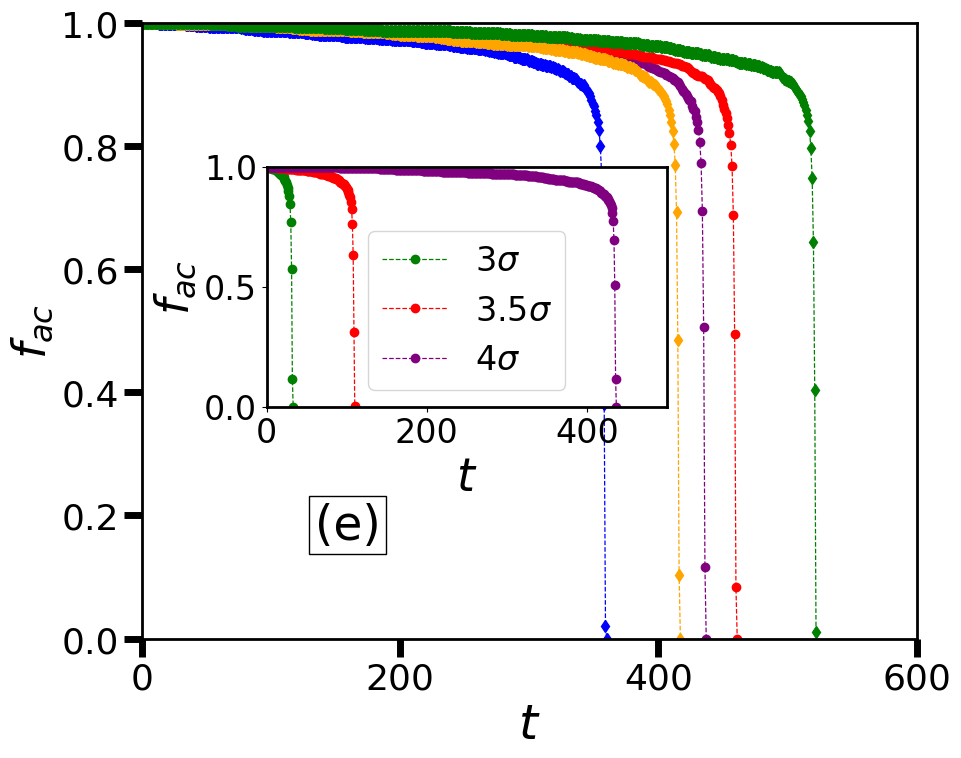}
\includegraphics*[width=0.32\textwidth]{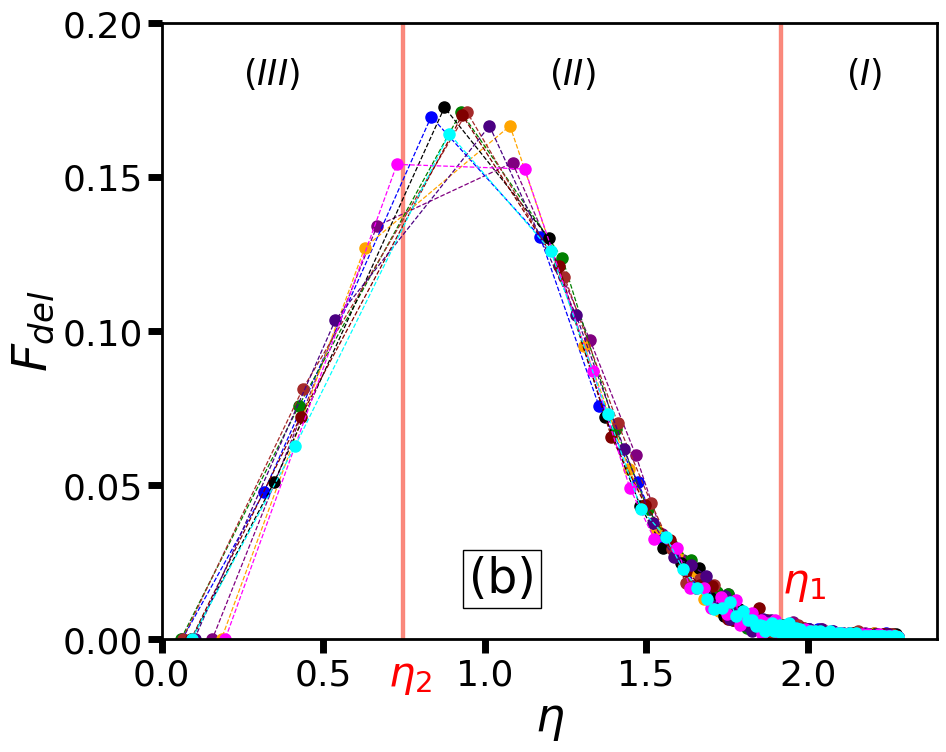}
\includegraphics*[width=0.32\textwidth]{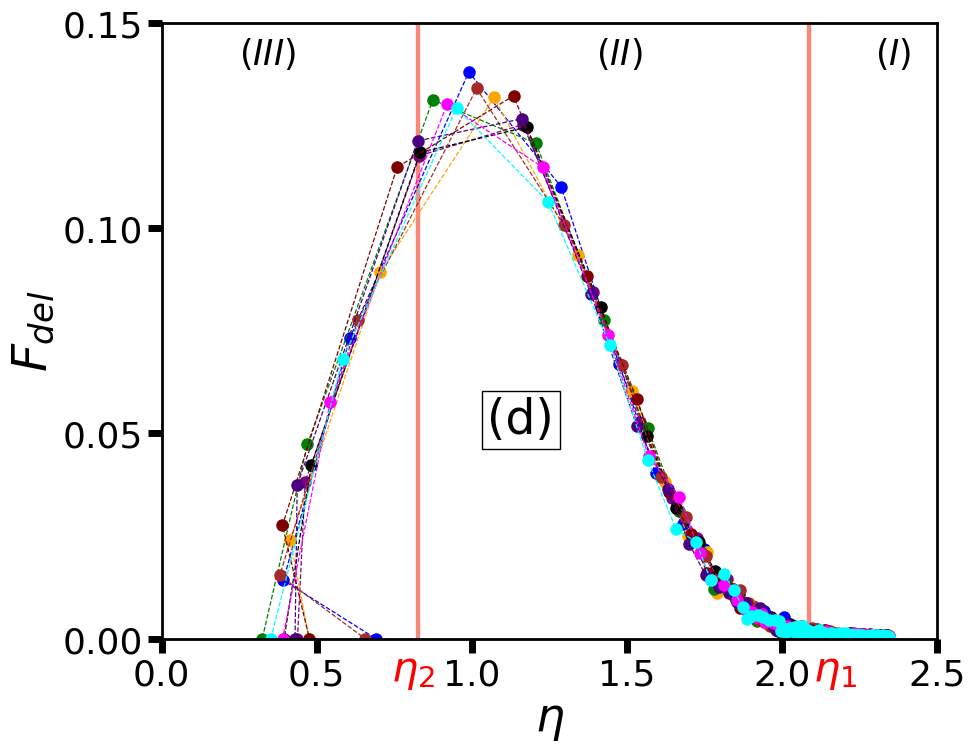}
\includegraphics*[width=0.32\textwidth]{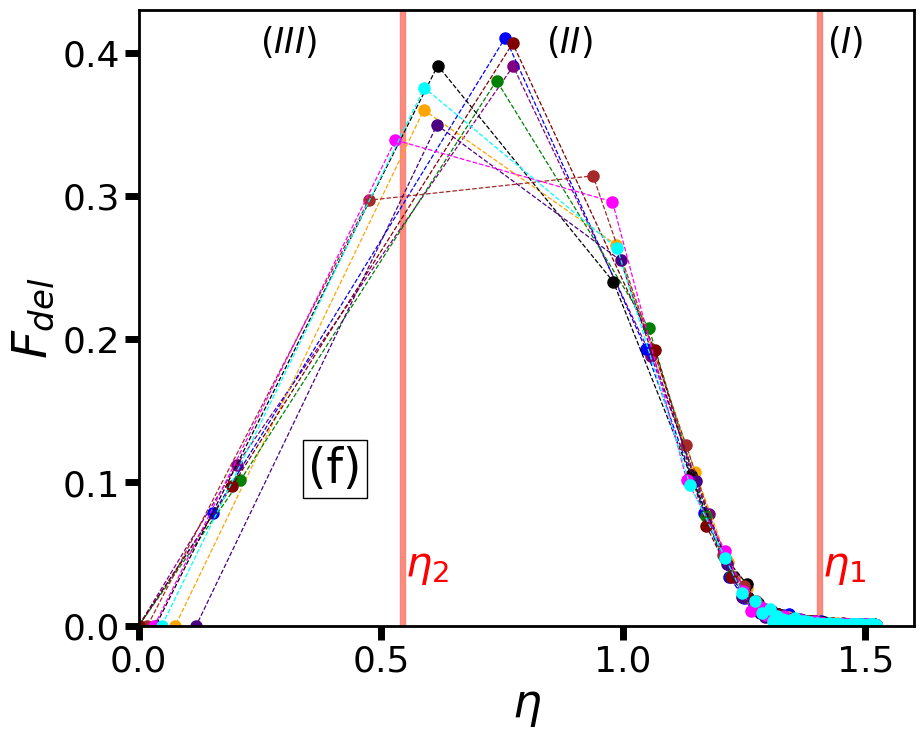}

\caption{Approach to network failure is depicted as a function of time and $\eta$. Fraction of active nodes as a function of time (top), and instantaneous fraction of deleted nodes as a function of $\eta$ (bottom) for $q=4$. Approach to failure due to extreme events on (a,b) Erdos-Renyi network with $\eta_{max} = 2.28$, (c,d) Barabasi-Albert network with $\eta_{max}=2.35$, and (e,f) small world network with $\eta_{max}=1.52$. The insets in (a,c,d) shows three cases for different threshold values. In (b,d,f), $\eta_1$ and $\eta_2$ demarcate the boundaries between the three regimes -- independent failure regime [indicated as (I)], cascade regime [indicated as (II)], and overload regime [indicated as (III)]. See text for details about how $\eta_1$ and $\eta_2$ are determined.}
\label{fig:appfail1}
\end{figure*}

When a node encounters an extreme event, it is disconnected from the network. Consequently, other active nodes (even if they did not encounter an extreme event) may also lose one or more edges. At time instants when node failure happens, degree $\kappa_i$ of $i$-th node changes while $W$ remain unchanged. The diameter of this network being unity, immediately after the first node failure random walkers are spread over the entire network. Based on this assumption and Eq. \ref{eq:eep1}, the probability $\phi_{{\rm del},i}$ of finding a walker on $i$-th node {\it after} $N_{\rm del}$ nodes have failed (and hence deleted from network) is
\begin{equation}
\phi_{{\rm del},i}(t) = \frac{\kappa_i(t)}{\sum_{j=1}^{N-N_{\rm del}(t)} \kappa_j(t) },
\label{eq:pdel1}
\end{equation}
where $\kappa_i$ is the degree of $i$-th node after $N_{\rm del}$ nodes are deleted. Time-dependence in Eq. \ref{eq:pdel1} arises because extreme events induce nodal failures and removal of edges during temporal evolution of walkers. For an all-to-all network, note that $\kappa_i=\kappa$ for all $i$. Hence, we suppress the index $i$. Thus, we have $\kappa_i(t) = \kappa(t) = N - 1 - N_{\rm del}(t)$. Using this in Eq. \ref{eq:pdel1}, we obtain a simplified form as
\begin{equation}
\phi_{\rm del}(t) = \frac{1}{N- N_{\rm del}(t)}.
\label{eq:pdel2}
\end{equation}
After $N_{\rm del}$ nodes are deleted, the probability for nodal failure (which equals the probability for the occurrence of extreme events) on any node can be obtained as
\begin{equation}
F({N}_{\rm del}(t)) = I_{\phi_{\rm del}(t)} \left( \floor*{\tau}+1, W-\floor*{\tau} \right).
\end{equation}
The fraction of deleted nodes can be expressed as
\begin{equation}
F_{\rm del}(t)= N_{\rm ac}(t) ~ F({N}_{\rm del}(t)).
\label{eq:delf1}
\end{equation}
Now, we will track the approach to network failure by eliminating time in $F_{\rm del}(t)$ and $\eta(t)$. Note that both $F_{\rm del}(t)$ and $\eta(t)$ change only at time instants when extreme event occurs anywhere on the network. Figure \ref{fig1}(b) displays $F_{\rm del}(t)$ (dotted line) as a function of $\eta(t)$. The simulation results from random walk realizations are shown as colored symbols. An excellent agreement between Eq. \ref{eq:delf1} and the simulation results are observed. This also allows us to clearly identify three distinct regimes as the complete network failure is approached on a timescale of $T_{\rm fail}$. These regimes are separated by vertical lines in Fig. \ref{fig1}(b), and are discussed in detail below.

{\it Independent node failures :} First of these is the regime of independent node failures in which failure of one node does not directly influence the other nodes to fail. Typically, at most, one node fails at a given time, and the time interval $\Delta t$ between two consecutive node failures is large; $\Delta t >> 1$. These imply that node failures tend to be rare, and do not disturb the network function.
Based on Eq. \ref{eq:eta}, let $\eta_{\rm 0}=\eta(t=0)$ be the maximum value of $\eta$ for a given network configuration. Then, this regime  begins at $\eta=\eta_{\rm 0}$, and breaks down at $\eta=\eta_{\rm 1}$ when the condition 
\begin{equation}
\Bigm| \left\langle \frac{dF_{\rm del}}{d\eta} \right\rangle_{\delta\eta} \Bigm| ~ > 0
\label{eq:netf1}
\end{equation}
is satisfied for the first time (see Fig. \ref{fig1}). In this, $\langle . \rangle$ represents an average over intervals $\delta \eta$, where $\delta \eta \ll \eta_{max}$. 
%In terms of time, it is  such that $1 < \delta t \ll T_{\rm fail}$. 
In Fig. \ref{fig1}(b), this condition is satisfied for $\eta = \eta_1$, and it is marked by a vertical line. The independent node failure regime lasts until $\eta > \eta_1$  and is indicated as Region I in Fig. \ref{fig1}(b).

{\it Cascade failures :} In this regime, as $\eta$ decreases below  $\eta_1$, the number of deleted nodes per time step is now more than 1, {\it i.e.}, $n_{del} > 1$. The node failures in this regime are not independent since every one node failure leads many other nodes to fail. In some parts of this regime, the total capacity is more than the total load present on the network ($\eta > 1$), but only some of the nodes encounter extreme events (and get disabled) due to excess load received from the redistribution of walkers. In an all-to-all network, all the nodes are neighbors of each other and hence, these failures are correlated with one another. Failure of a small number of nodes is followed by the failure of a larger number of nodes at the next time step. Such a sequence of node failures is called the cascade failures, and this regime lasts until $\eta=\eta_2$, where $\eta_2$ is determined through 
\begin{equation}
\eta_2 = \alpha ~ {\rm argmax}_{\eta} \Bigm| \left\langle \frac{dF_{\rm del}}{d\eta} \right\rangle_{\delta\eta} \Bigm|. 
\label{eq:netf1}
\end{equation}
%{\color{blue}$\eta_2 = \alpha ~ \eta_{\rm max}$, $\eta_{\rm max} = \Bigm| \left\langle \frac{dF_{\rm del}}{d\eta} \right\rangle \Bigm| ~ = 0$, calculated when $\eta < \eta1$.}
In this, $\alpha$ is a parameter to be chosen based on empirical simulations. Our simulations show that choosing $\alpha=1$ underestimates $\eta_2$, while $\alpha=0.8$ provides a more realistic estimate for $\eta_2$ associated with regime change. The cascade failure regime is shown as Region II in Fig. \ref{fig1}(b). 
%The transition from independent to cascade failure regime becomes sharper as the threshold $\tau$ is increased. 

{\it Overload failure :} In this regime, the entire network fails with deletion of all the remaining nodes in one time step, or in no more than 2 time steps. In this case, as it is a one-step failure, number of deleted nodes equals the instantaneous number of active nodes, and total load on the network is significantly greater than its carrying capacity, {\it i.e.}, $\eta << 1$. As a consequence, each node is already overloaded, and typically it takes no more than one time step for all the nodes to fail together. This is evident for $\eta < \eta_2$ (shown as Region III in Fig. \ref{fig1}(b)), since for each realization of the random walk dynamics there is just one data point appearing in this regime.

Compared to the temporal picture shown in Fig. \ref{fig1}(a), an elegant characterization of the approach towards network failure -- triggered by the intrinsic fluctuations in flux -- emerges when the same data is examined as a function of $\eta$ as displayed in \ref{fig1}(b). In particular, the analytical result in Eq. \ref{eq:delf1} provides a good description of the simulation results and helps to identify three distinct regimes. 
The $\eta_{1}$ is obtained by taking the running averages of the multiple realizations in a particular network and estimating the change in slope over a period of capacity-to-load variation (This varies with the types of networks). The $\eta_{2}$ is estimated considering $0.8$ times the peak fraction of deletion experienced by a particular network. These values give a good overview of the phases of the network that can also be visualized for synthetic and real-world networks. The next natural question is to ask if these three regimes occur in other types of network as well. We address this question below.

In this subsection, we examine the approach to network failure in Erdos-Renyi (ER), Barabasi-Albert (BA), and small world (SW) networks. The simulation results for all the three types of networks are depicted in Fig. \ref{fig:appfail1}. These networks are generated using {\tt NetworkX \cite{networkx_11}}. We employ an Erdos-Renyi network with $N=2000$ nodes and $E=10028$ edges and diameter $d_{er}=6$, a small-world network with $N=2000$ and $E=60000$ and $d_{sw}=4$, Barabasi-Albert network with $N=5000$ and $E=19915$ and $d_{ba}=6$, and lastly a square lattice. Random walk simulations for $T$ timesteps are performed with $W=2~E$ walkers whose initial positions at time $t=0$ are randomly chosen. Before extreme events are monitored, $5000$ time steps of random walk are ignored to ensure that steady state is reached.

The broad features of the approach to network failure as a function of time (Fig. \ref{fig:appfail1}(a-c)) and with $\eta$ (Fig. \ref{fig:appfail1}(d-f)) are qualitatively similar to that observed in all-to-all network in Fig. \ref{fig1}. Hence, it can be stated that approach to network failure -- induced by intrinsically generated extreme events -- is qualitatively similar for ER, BA and SW networks. In particular, the three regimes identified and described for all-in-all case hold good in Fig. \ref{fig:appfail1}(d-f) as well. The vertical lines shown in Figs. \ref{fig:appfail1}(d-f) demarcate the independent, cascade and overload regimes of network failure. Further, as $t \to \infty$, the size of the largest connected component $L$ (not shown here) tends to zero or a small number indicating that network has effectively lost its functional capability to process flux.

Despite the broad similarity of the features in Fig. \ref{fig:appfail1} to that in Fig. \ref{fig1}, there are differences arising from the nature of connectivity in the  networks other than all to all. For instance, in these networks, as $t \to \infty$, nodes get isolated from the bulk of the network with walkers trapped in those nodes. The fraction of nodes getting isolated may differ in different realizations and different networks. Isolated nodes are essentially inactive nodes, and the $W_{\rm trap}$ trapped walkers on them cannot perform any dynamics. The effective number of walkers is $ W_{\rm eff} = W - W_{\rm trap}$. This situation can arise even if a node does not encounter extreme events. Hence, due to this, both capacity and load are decreasing but rate of decrease differ depending upon the network topology. For instance, in BA network, the rate of decrease in load $W_{\rm eff}$ is faster than that of capacity $C$. Thus, the ratio $\eta(t)=C(t)/W_{\rm eff}$ increases, and consequently the network shows an inclination to return to the independent regime.  This effect is visible in Fig. \ref{fig:appfail1}(d), and this becomes more pronounced as threshold $\tau$ increases. Further, maximum fraction of nodes deleted at a single step depends on the diameter of the network -- shorter the diameter more number of nodes are deleted. Hence, it  is highest in an all-to-all network. This can be inferred by comparing the peak values in Fig. \ref{fig:appfail1}(b,d,f).

\begin{figure}[t]
\includegraphics*[width=0.65\linewidth]{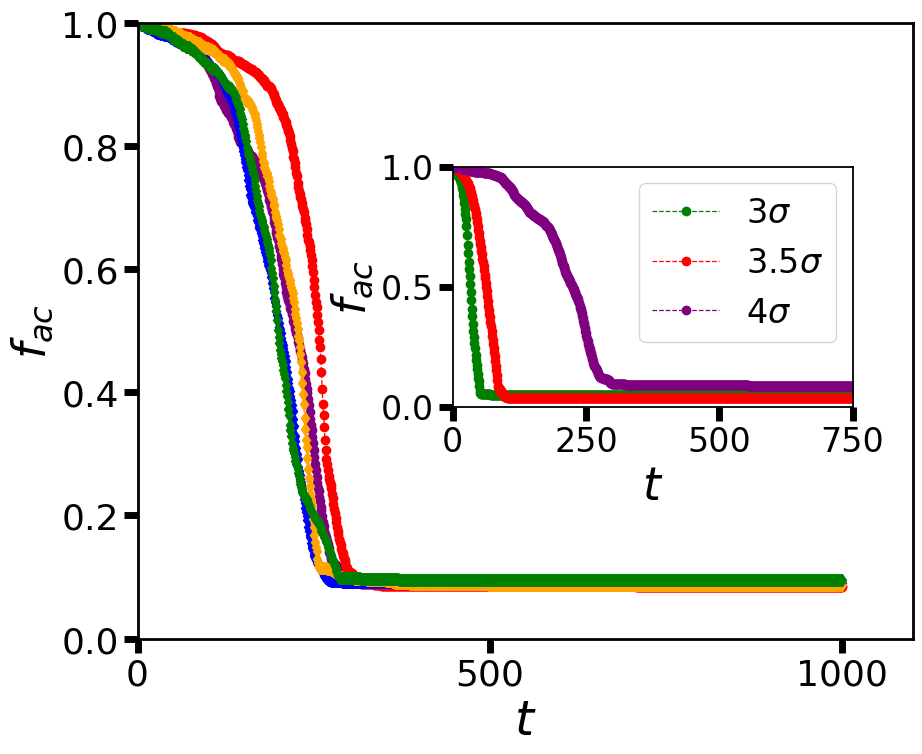}
\includegraphics*[width=0.65\linewidth]{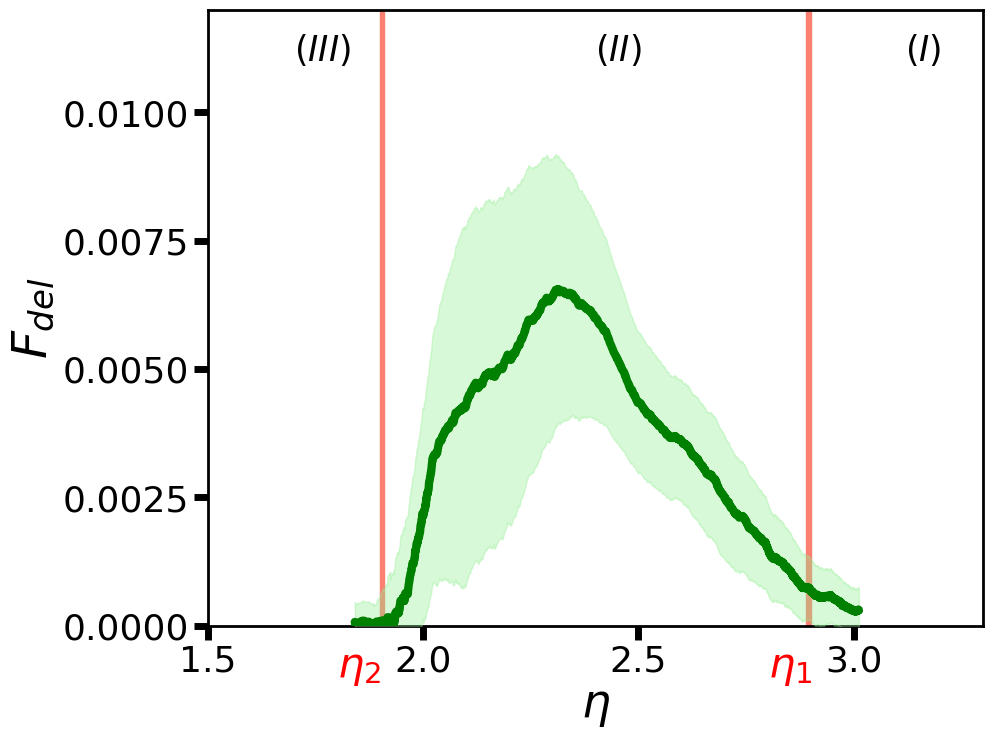}
\caption{Approach to network failure on a square lattice. (top) depicts the fraction of active nodes as a function of time, (bottom) shows the fraction of deleted nodes as a function of $\eta$ for $q=4$. $\eta_1$ and $\eta_2$ represent boundaries between the regimes and $\eta_{max}=3.02$. The solid line is the running average $\langle F_{\mathrm{del}}(\eta) \rangle$ over 10 initial conditions. 
The light green band shows the standard deviation $\sigma_{F_{del}}(\eta)$ within  a fixed window of $\Delta\eta = 0.1$.}
\label{fig:appfail2}
\end{figure}

Next, we consider a square lattice network with $N=2500$ nodes and $E=4900$ edges. The random walk simulations were performed with $W=2~E$ walkers and the results are displayed in Fig. \ref{fig:appfail2}. The decay of the fraction of active nodes $f_{ac}$ as a function of time is shown in Fig. \ref{fig:appfail2}(top) for a few realizations. Unlike similar results shown for the case of heterogeneous networks in Fig. \ref{fig:appfail1}, $f_{ac}$ displays nearly the same decay profile in every realization. Further, notice that $f_{ac}$ never reaches zero but saturates at $f_{ac} \approx 0.15$. This happens because, as $t \to \infty$, a very small number of connected nodes, often in different clusters, tend to survive and support dynamics though the network as a whole has effectively failed and cannot support dynamics.

Figure \ref{fig:appfail2}{(bottom)} shows the fraction of inactive nodes $F_{\rm del}$ as a function of $\eta$.

As in other heterogeneous networks discussed earlier, the three failure regimes can still be identified; however, the underlying topology of the square lattice leads to important qualitative differences. In a square lattice, node failures are not strictly independent, since the absence of long-range connections and the resulting large network diameter confine the redistribution of load to local neighborhoods. Each node has a maximum degree of four, and therefore the failure of a node is accompanied by the removal of at most four edges. The load carried by the failed node is consequently transferred only to its immediate neighbors, which are themselves topographically adjacent. This local redistribution leads to a progressive acceleration in the rate of node deletion as failures spread across the lattice. The deletion rate reaches a maximum when the cascade spans a large fraction of the network and boundary nodes also begin to fail. In contrast to well-connected networks studied earlier, where long-range connectivity enables rapid global load sharing, the square lattice lacks such pathways. As a result, walkers accumulate locally, leading to a rapid build-up of load in confined regions. This enhances the probability of extreme events and drives the system into the cascade regime. However, unlike in well-connected networks, these cascades remain spatially localized and are comparatively smaller in extent. As the dynamics progresses further, the rate of deletion slows down due to extensive fragmentation of the network. A large number of isolated nodes and small disconnected clusters are formed, each supporting only a limited amount of load. Consequently, there is no distinct global overload failure regime. Instead, residual activity persists within these disconnected components until the end of the simulation. Although the square lattice loses most of its nodes and edges, a few disconnected clusters survive for very long times and may never fail. At this stage, the network has effectively ceased to support meaningful transport or collective dynamics.
\begin{figure}[t]
\includegraphics*[width=0.55\linewidth]{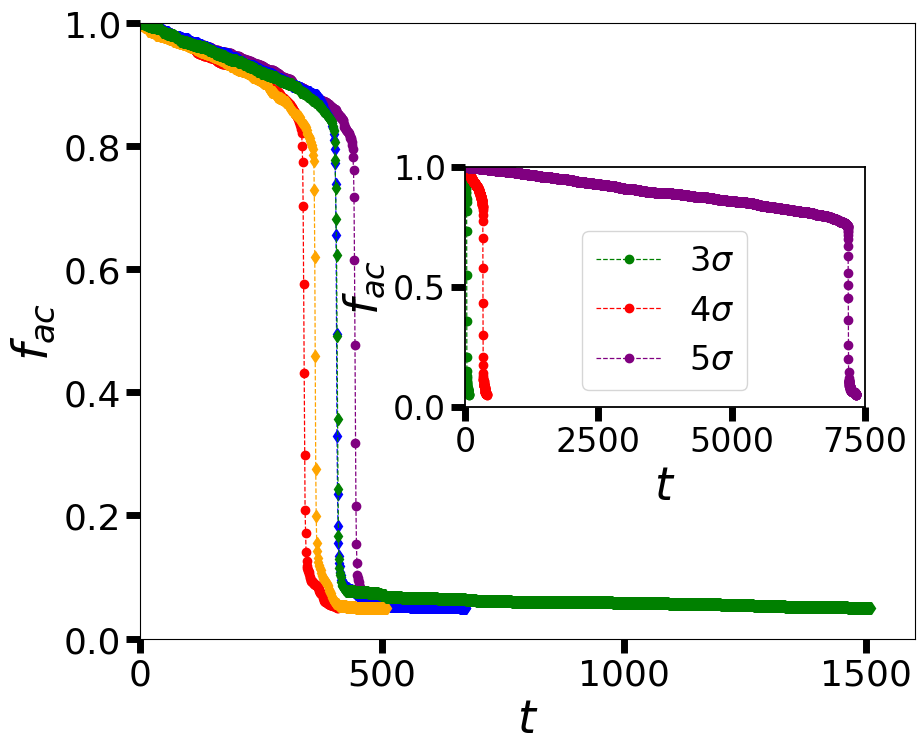}
\includegraphics*[width=0.55\linewidth]{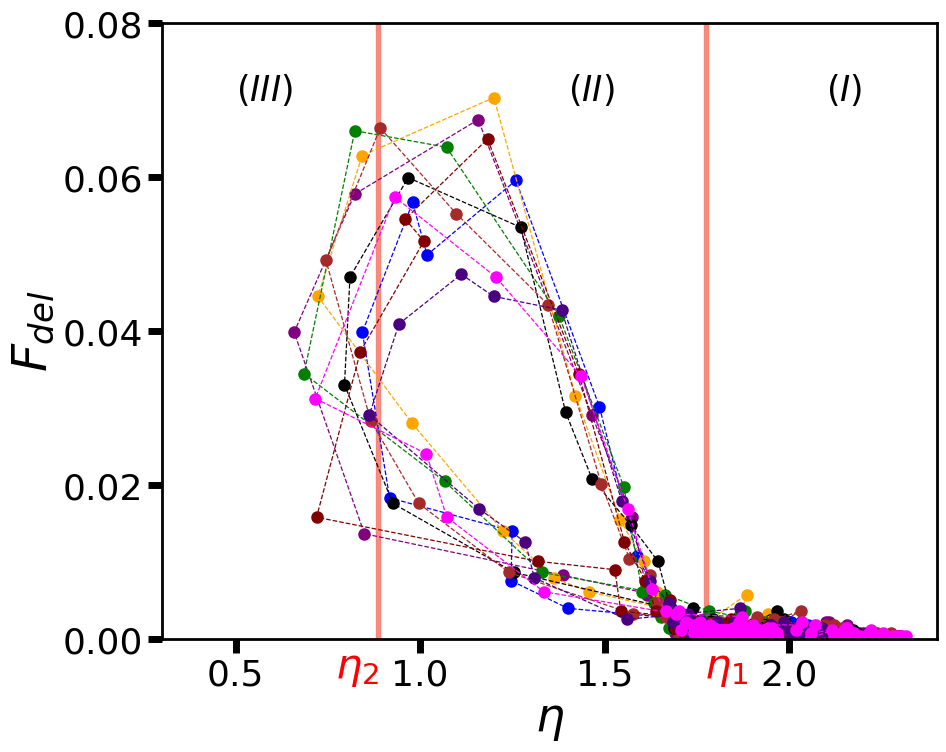}
\caption{Approach to network failure for a flight network. (top) the fraction of active nodes as function of time. (bottom) the fraction of deleted nodes against $\eta$. $\eta_1$ and $\eta_2$ represent boundaries between the three regimes, also $\eta_{max}=1.96$. Due to significant fraction of trapped walkers for $\eta < 1$, overload regime is marked by a strong tendency to return to independent regime.}
\label{fig:appfail3}
\end{figure}

We apply the idea of network failure to a flight network with $N=2789$ nodes, $E=14795$ edges, and diameter $d_{\rm fl}=14$. The dataset was obtained from the Network Repository~\cite{nr}. This network is of ``Barabasi-Albert'' type, and  differs from synthetic networks due to its large diameter $d_{\rm fl}=14$, compared to $d_{\rm ba} = 6$ for a Barabasi-Albert network of similar size. As before, we obtain the decay of active nodes $f_{ac}$ as a function of time for a few realizations. Here, the decay profile (Fig. \ref{fig:appfail3}(top)) is similar to that of a scale free network, but a new steady state forms after the "cascading" failure regime where most of the network has failed. However, due to the formation of disconnected segments the dynamics continues for a considerable amount of time. The network fragments, and has isolated segments which continue to support dynamics. Hence, due to trapped walkers (similar to scale-free network), a return to the independent failure regime is observed for $t \gg 1$ in Fig. \ref{fig:appfail3}(bottom).

%\begin{table}[h!]
%\centering
%\caption{Network regions}
%\begin{tabular}{|c|c|c|c|}
%\hline
% Network & $eta1$ & Peak & $eta2$ \\
% \hline
% All to all & 1.23 & 0.88 & 0.70 \\
% Erdos-Renyii & 1.92 & 0.94 & 0.75 \\
% Small world & 1.40 & 0.69 & 0.55 \\
% Barabasi-Albert & 2.09 & 1.04 & 0.83 \\
% Square lattice & 2.9 & 2.33 & 1.9 \\
% Flight & 1.78 & 1.12 & 0.89 \\
% \hline
% \end{tabular}
% \end{table}

In conclusion, we studied the network failures induced by large intrinsic fluctuations in the flux of random walkers modelled as extreme events, devoid of any external stimulus. While network failures from external perturbations are well studied \cite{AlbAlbNak2004,PagAie2013,gajduk2014stability}, we demonstrate that intrinsic fluctuations alone can lead to network collapse. Crucially, failure proceeds through three regimes: (i) uncorrelated failures, (ii) cascading collapses, and (iii) overload (one-step failure) regime. These regimes emerge across diverse networks and are analytically described for all-to-all networks (Eq.~\ref{eq:delf1}), matching well with simulations.
The results challenge the paradigm that resilience requires only protection against external shocks. Instead, internal fluctuations -- ubiquitous in power grids, traffic, and communication systems -- require new mitigation strategies. While we assumed permanently failed nodes, real systems often allow recovery. Future work could extend this framework by introducing repair dynamics to explore resilience in more realistic settings.

\begin{acknowledgments}

The authors acknowledge the use of computational resources provided by the PARAM Brahma supercomputing facility under the National Supercomputing Mission (NSM) at IISER Pune. The facility is implemented by C-DAC and supported by the Ministry of Electronics and Information Technology (MeitY) and the Department of Science and Technology (DST), Government of India.

\end{acknowledgments}

\vspace{2em}
\noindent
\sloppy
\nocite{*}
\bibliographystyle{apsrev4-2}
\bibliography{netbiblio}

\end{document}